\documentstyle[12pt]{article}
\begin{document}

\begin{center}
{\bf
ANALYTIC FORM FACTORS OF HYDROGENLIKE ATOMS\\
    FOR DISCRETE-CONTINUUM TRANSITIONS.\\
    I. TRANSITIONS FROM $nS$-STATES \\}
\vspace*{.5cm}
O.Voskresenskaya$^{1,2,3}$\footnote{E-mail:
Olga.Voskresenskaja@mpi-hd.mpg.de; voskr@cv.jinr.dubna.su}\\
\vspace*{.25cm}
{\it$^1$ Institut f\"ur Theoretische Physik der Universit\"at,
69120 Heidelberg, Germany\\
\it$^2$ Max-Planck-Institut f\"ur Kernphysik, Postfach 103980,
69029 Heidelberg, Germany\\
$^3$ Joint Institute for Nuclear Research,
141980 Dubna, Moscow Region, Russia\\}
\end{center}
\vspace*{.25cm}
\begin{abstract}
{\small The new approach for calculation of transition form factors
of hydrogenlike atoms is proposed. The explicit expressions for form
factors of transitions from bound $nS$-states to continuum in terms of
the classical polynomials are derived}

\end{abstract}
\vspace*{.5cm}

The calculation of transition form factors from bound to unbound (continuum)
states of hydrogenlike atoms have been discussed in series of papers
\cite{omid}. Nowadays, this problem became of great importance due to 
preparing of the DIRAC experiment at CERN \cite{dirac,dubna,HadAt99,HadAt01}.

The usual approach to this problem is based on the decomposition of continuum
wave functions into infinite series of partial waves and calculation of
transition form factors from initial bound state to final continuum state with
definite value of angular momentum.

In this approach only finite number of terms of infinite series are taken into
account in actual calculation leaving unsolved the problem of 
estimation for contribution of omitted tail with infinite number of 
terms.

We would like to show in this paper that above mentioned transition (bound to
continuum) form factors of hydrogenlike atoms may be explicitly calculated
without decomposition of final state into infinite series of partial waves.

Below we shall prove this statement for the simplest case when orbital
momentum of bound state is equal zero, i.e. we restrict of ourselves
with initial $nS$-states.
The generalization of this consideration for the case of arbitrary initial
states will be done later.

We define the transition form factors with the help of the following 
equation:  \begin{equation} S_{fi}(\vec q)=\int \psi_f^{\ast}(\vec 
r)e^{i\vec q\vec r} \psi_i(\vec r)d^3r \label{e1}\, , \end{equation} 
where $\psi_{i(f)}$ are the wave functions of initial (final) states
of hydrogenlike atoms, $\vec q$ is the transferred momentum.

For the case when  $i=n00\equiv nS$
\begin{eqnarray}
\psi_i=\psi_{n00}(\vec r)&=&\left(\frac{\omega^3}{\pi}\right)^{\frac{1}{2}}
\exp(-\omega r)\cdot\Phi(-n+1;2;2\omega r)\nonumber\\
&\equiv&\left(\frac{\omega^3}{\pi}\right)^{\frac{1}{2}}n^{-1}
\exp(-\omega r)\cdot L^{1}_{n-1}(2\omega r)
\label{e2}\, ,
\end{eqnarray}
where $\omega=\mu\alpha/n$; $\mu$ is the reduced mass and 
$\alpha=1/137$ is the fine structure constant; $\Phi$ is the confluent 
hypergeometric function and $L^{\lambda}_k$ are the associated Laguerre 
polynomials.

The wave function of the final (continuum) state must be choose in the form
\cite{land}
\begin{eqnarray}
\psi_f(\vec r)=\psi_{\vec p}^{(-)}&=&c^{(-)}\exp(i\vec p\vec r)
\cdot\Phi\left[-i\xi,1,-i(pr+\vec p\vec r)\right]\,,
\label{e3}
\end{eqnarray}
$$c^{(-)}=(2\pi)^{-\frac{3}{2}}\exp\left(\frac{\pi\xi}{2}\right)
\Gamma(1+i\xi)\,,$$
$$\xi=\frac{\mu\alpha}{p}\,.$$

Now we use recurrence relation for the Laguerre polynomials \cite{ryzh}
\begin{equation}
L_k^{\lambda+1}(x)=\frac{1}{x}\left[(k+\lambda+1)L_{k-1}^{\lambda}(x)
-(k+1)L_{k}^{\lambda}(x)\right]
\label{e4}
\end{equation}
and their representation with the help of the generating function
\begin{equation}
L_k^{\lambda}(x)=\Delta_z^{(k)}\left[(1-z)^{-(\lambda+1)}\exp
\left(\frac{xz}{z-1}\right)\right]
\label{e5}\, ,
\end{equation}
where operator $\Delta_z^{(k)}$ is defined as follows:
\begin{equation}
\Delta_z^{(k)}\left[f(z)\right]=\frac{1}{k!}
\left.\left(\frac{d^k}{dz^k}f(z)\right)\right\vert_{z=0}
\label{e6}\, .
\end{equation}

Then
\begin{equation}
\psi_i(r)=\frac{1}{2r}\left(\frac{\omega}{\pi}\right)^{\frac{1}{2}}
\left[\Delta_z^{(n-1)}-\Delta_z^{(n)}\right]
\left[(1-z)^{-1}\exp\left[-\omega(z)r\right]\right]
\label{e7}\, ,
\end{equation}
\begin{equation}
\omega(z)=\omega\cdot \frac{1+z}{1-z}\,.
\label{e11}
\end{equation}

The substitution of Eqs. (\ref{e3}) and (\ref{e7}) to (\ref{e1}) gives
\begin{equation}
S_{\vec p,noo}(\vec q)=\frac{1}{2}\left(\frac{\omega}{\pi}\right)^{\frac{1}{2}}
c^{(-)}
\left[\Delta_z^{(n-1)}-\Delta_z^{(n)}\right]
\left[\frac{J(\vec q,\vec p,z)}{(1-z)}\right]
\label{e8}\, ,
\end{equation}
\begin{equation}
J(\vec q,\vec p,z)=\int \frac{d^3r}{r}
\Phi\left[i\xi,1,i(pr+\vec p\vec r)\right]
\exp[i(\vec q-\vec p)\vec r-\omega(z) r]\,.
\label{e9}
\end{equation}

The last integral is easily calculated  using integral representation 
for the hypergeometrical functions (see e.g. \cite{nord}).

The result reads
\begin{equation}
J(\vec q,\vec p,z)=4\pi[\omega^2(z)+\vec\Delta^2]^{-1+i\xi}
\left[[\omega(z)-ip]^2+q^2\right]^{-i\xi}\,,
\label{e10}
\end{equation}
where $\vec\Delta=\vec q-\vec p$.

Taking into account the definition (\ref{e11})
and obvious the relation
\begin{equation}
\Delta_z^{(n)}\left[zf(z)\right]=\Delta_z^{(n-1)}f(z)\,,
\label{e12}
\end{equation}
(\ref{e8}) may be rewritten in the form
\begin{equation}
S_{\vec p,noo}(\vec q)=-4(\pi\cdot\omega)^{\frac{1}{2}}c^{(-)}
\left(\Delta_z^{(n)}-2\Delta_z^{(n-1)}+\Delta_z^{(n-2)}\right)
\left(D_1^{-1+i\xi}D_2^{-i\xi}\right)
\label{e13}\, ,
\end{equation}
$$D_1=(1+z^2)(\omega^2+\vec\Delta^2)-2z(\vec\Delta^2-\omega^2)\,,$$
$$D_2=(\omega-ip)^2+q^2-2z(q^2-p^2-\omega^2)+
z^2\left((\omega+ip)^2+q^2\right)\,.$$

Using the definition of the Gegenbauer polynomials \cite{ryzh,abram}
\begin{equation}
(1-2xz+z)^{-\lambda}=\sum_{k=0}^{\infty}C_k^{(\lambda)}(x)\cdot z^k\,,
\label{e14}
\end{equation}
it is easy to obtain
\begin{equation}
D_1^{-1+i\xi}=(\Delta^2+\omega^2)^{-1+i\xi}
\sum_{k=0}^{\infty}C_k^{(1-i\xi)}(u)\cdot z^k
\label{e15}\,,
\end{equation}
$$u=\frac{\Delta^2-\omega^2}{\Delta^2+\omega^2}\,;$$
\begin{equation}
D_2^{-i\xi}=[(\omega-ip)^2+q^2]^{-i\xi}
\sum_{k=0}^{\infty}C_k^{(i\xi)}(v)\cdot w^k\cdot z^k
\label{e16}\,,
\end{equation}
$$v=\frac{q^2-p^2-\omega^2}{\sqrt{[(\omega-ip)^2+q^2][(\omega+ip)^2+q^2]}}\,,$$
$$w=\sqrt{\frac{(\omega+ip)^2+q^2}{(\omega-ip)^2+q^2}}\,.$$

At least, with the help of the relations
\begin{equation}
\Delta_z^{(n)}[f_1(z)f_2(z)]
=\sum_{k=0}^{n}
\left[\Delta_z^{(n-k)}f_1(z)\right]\left[\Delta_z^{(k)}f_2(z)\right]
\label{e17}
\end{equation}
and
\begin{equation}
C_n^{(\lambda)}(x)-C_{n-1}^{(\lambda)}(x)=\frac
{\Gamma(\lambda-\frac{1}{2})\Gamma(n-1+2\lambda)}
{\Gamma(2\lambda-1)\Gamma(n+\lambda-\frac{1}{2})}
\cdot P_n^{(\lambda-\frac{3}{2},\lambda-\frac{1}{2})}(x)\,,
\label{e18}
\end{equation}
where $P_n^{(\lambda-\frac{3}{2},\lambda-\frac{1}{2})}(x)$ are the Jacobi
polynomials, we finally obtain
\vspace*{.25cm}
\begin{equation}
S_{\vec p,noo}(\vec q)=-2(\pi\cdot\omega)^{\frac{1}{2}}\cdot c^{(-)}
\cdot(\omega^2+\Delta^2)^{-1+i\xi}
\label{e19}\end{equation}
$$\times[(\omega-ip)^2+q^2]^{-i\xi}
\frac{\Gamma\left(\frac{1}{2}- i\xi\right)}
{\Gamma\left(1- 2i\xi\right)}\cdot
\sum_{k=0}^{n}w^kC_k^{(i\xi)}(v)$$

$$
\times\left[\frac{\Gamma\left(n-k+1-2i\xi\right)}{\Gamma\left(n-k+\frac{1}{2}- i\xi\right)}\cdot
P_{n-k}^{(-\frac{1}{2}-i\xi,\frac{1}{2}-i\xi)}(u)
-\frac{\Gamma\left(n-k- 2i\xi\right)}{\Gamma\left(n-k-\frac{1}{2}- i\xi\right)}\cdot
P_{n-k-1}^{(-\frac{1}{2}-i\xi,\frac{1}{2}-i\xi)}(u)\right].
$$
\vspace*{.5cm}
\newpage

Thus, the result of the calculation for transition form factors
of  hydrogenlike atoms for the case  $nS$-continuum transitions
is expressed in terms of the classical polynomials and may be
easily evaluated numerically with arbitrary degree of accuracy.
\vspace*{1cm}

\hspace{-.5cm}{\large\bf Acknowledgments}
\vspace*{1cm}

The author is grateful to Professor J\"org H\"ufner for invitation and
hospitality at the Institute for Theoretical Physics, Heidelberg University,
where this work was done, and to Alexander Tarasov for useful discussions.

\end{document}